\documentclass[twocolumn,showpacs,preprintnumbers,amsmath,amssymb]{revtex4}
\usepackage{tabularx,graphicx}\begin{document}
%\documentstyle[aps]{revtex}
%\documentstyle[preprint,aps]{revtex}
%\begin{document}
\newcommand{\beq}{\begin{equation}}
\newcommand{\eeq}{\end{equation}}
\newcommand{\beqn}{\begin{eqnarray}}
\newcommand{\eeqn}{\end{eqnarray}}
\newcommand{\bmath}{\begin{subequations}}
\newcommand{\emath}{\end{subequations}}
%\draft
\title{Apparent increase in the thickness of superconducting particles at low temperatures measured by electron holography
}
\author{J. E. Hirsch }
\address{Department of Physics, University of California, San Diego\\
La Jolla, CA 92093-0319}
 
\date{\today} 
\begin{abstract} 

We predict that superconducting particles will show an apparent increase in thickness at low temperatures when measured by electron
holography. This will result not from a  real thickness increase, rather from an increase in the mean inner  potential sensed by the electron wave
traveling through the particle, originating in expansion of the electronic wavefunction and resulting negative charge expulsion from the interior to the surface of the superconductor,
 giving rise to an increase in the phase shift of the   electron wavefront going
through the sample relative to the wavefront going through vacuum.
The temperature dependence of the observed phase shifts will yield valuable new information on the physics of the superconducting state of metals.

\end{abstract}
\pacs{}
\maketitle

\begin{figure} [h]
\resizebox{7.5cm}{!}{\includegraphics[width=7cm]{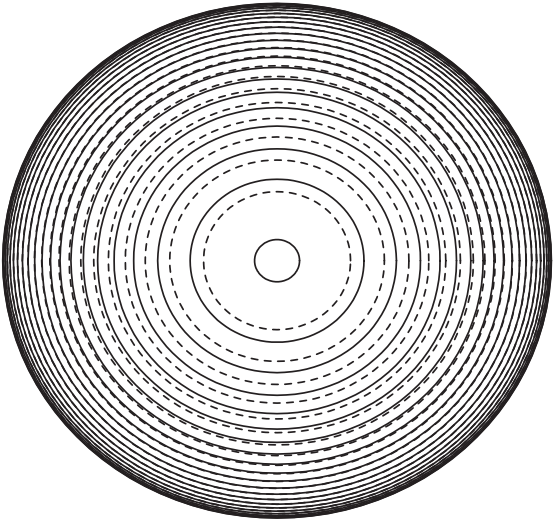}}
  \caption{Predicted interference micrograph of a spherical superconducting particle of radius $500nm$, mean inner potential $V_0=10V$, London penetration depth $\lambda_L=40nm$ and lower 
  critical magnetic field $H_{c1}=100G$ with $300kV$ electrons, two-times phase amplified. The full lines indicate the contours of constant phase above $T_c$, the dashed
  lines at temperatures well below $T_c$. }
\end{figure}

\begin{figure}
\resizebox{7.5cm}{!}{\includegraphics[width=7cm]{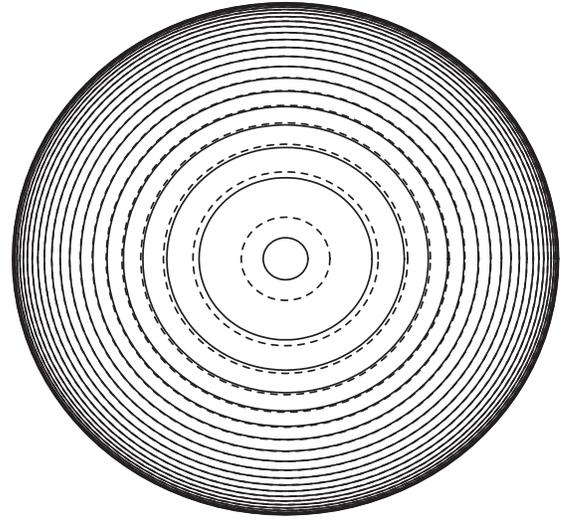}}
  \caption{ Same as Fig. 1 for $\lambda_L=150nm$ and $H_{c1}=73G$, parameters appropriate for $YBCO$.
  Both here and in Fig. 1, each full line contour is  predicted to move gradually $outward$ to the closest dashed line contour as the temperature is lowered below $T_c$.
  Phase amplification factor   is  $2$ as in Fig. 1.}
\end{figure}

\begin{figure}
\resizebox{5.5cm}{!}{\includegraphics[width=7cm]{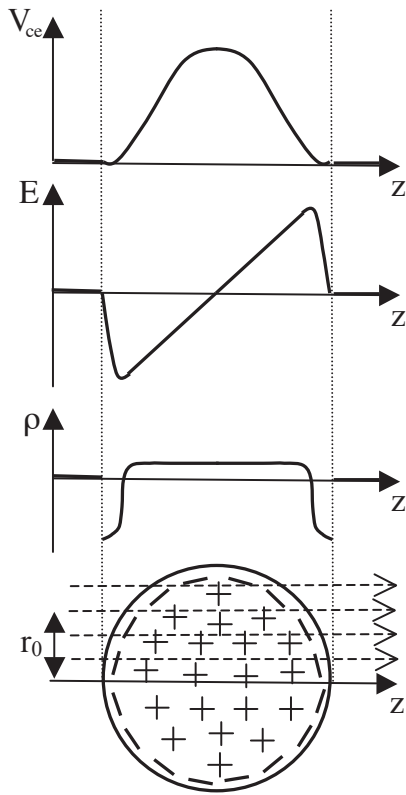}}
  \caption{ Predicted charge distribution for a superconducting spherical particle at low temperatures, and resulting charge density $\rho$, electric field $E$ and
  electric potential $V_{ce}$ (schematic) sensed by the beam electrons traveling along the sphere diameter. The electric field points radially outward. The charge density $\rho$ becomes negative at a distance $\lambda_L$
  (London penetration depth) from the
  surface, at that point the electric field magnitude reaches maximum value. The dashed lines show other possible trajectories for the beam electrons,
  all giving rise to a positive phase shift.}
\end{figure}

\section{introduction}

Electron holography is a sensitive tool to measure the thickness of small particles\cite{tonomura}, as demonstrated experimentally already several decades ago. Initially, experiments
yielded a resolution of only tens of nanometers\cite{endo} corresponding to phase difference of   $2\pi$ 
between neighboring contour lines. 
However it was soon shown that the resolution could be  improved by large factors using phase difference
amplification techniques, and phase shift differences of $2\pi/100$ could be detected\cite{endo2}.  Interference micrographs
such as shown in Figs. 1 and 2 depict  contours of constant phase shift, and for a spherical sample the contours are
circles as shown in those figures.  The phase difference between neighboring contours
in Figs. 1 and 2 is $\pi$ (two-times phase 
difference amplification).

A given circle of radius $r$ in the contour map corresponds to a given phase shift $\Delta \varphi$. Imagine the thickness of the sample were to increase as the temperature is lowered. The phase shift for that 
value of $r$ will increase, so the radius of the circle that will give to the original phase shift $\Delta \varphi$ has to increase. In other words, the contours
of given phase shift will move $outward$ as shown in Figs. 1 and 2 by the dashed lines. In this paper we predict that this will be observed for any superconducting particle cooled 
sufficiently below
its critical temperature. 

The contour shift will appear to indicate that the sphere becomes a prolate ellipsoid of revolution with the longer axis along the beam direction.
In reality, no change in the physical dimensions of the sample will have occurred. The behavior shown in Figs. 1 and 2 will signal that a 
change in the mean inner potential sensed by the electron beam will have occurred, originating in electronic charge redistribution inside the superconductor: namely, that negative charge has moved from the interior of the superconductor to the
surface. The predicted  charge distribution, electric field and electric potential inside the superconductor  is shown schematically in Figure 3.
For a spherical particle no electric field outside the particle is generated through this charge redistribution 
due to the spherical symmetry and the fact that   the particle remains 
charge neutral. 

 \begin{figure}
\resizebox{7.5cm}{!}{\includegraphics[width=7cm]{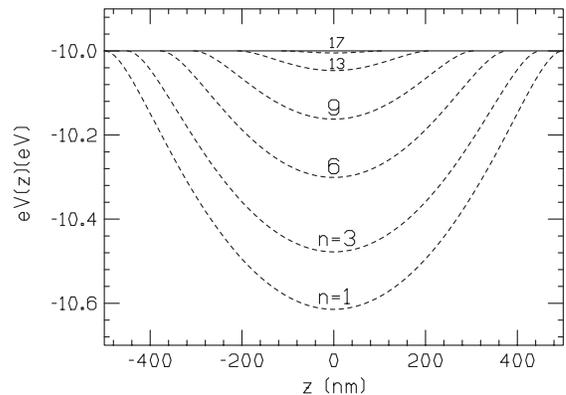}}
  \caption{ Electric potential energy of the electron wave going through the superconducting sphere, parameters as in Fig. 1. The labels $n$ correspond to
  the full contours in Fig. 1 starting with $n=1$ for the smallest contour. The full line at $eV=-10eV$ is the potential energy of the electron when the
  sample is in the normal state (mean inner potential $V_0=10V$). When the sample goes superconducting the potential energy drops (dashed lines) and the contours
  of given phase shift move outward as shown in Fig. 1.}
\end{figure} 

\section{phase shift}

In an electron holography experiment, the phase shift of the electron wave going through the particle at a distance 
 $r_0$ from the center of the sphere, relative to a reference wave going through vacuum is given by\cite{tonomura}
\beq
\varphi(r)=C_T\int_{-z_0(r_0)}^{z_0(r_0)} [V_0+V_{ce}(r(z,r_0))
\eeq
where
\beq
C_T=\frac{2\pi e}{\lambda T}\frac{T+E_0}{T+2E_0}
\eeq
with $T$ the kinetic energy of the electron, $E_0$ the electron rest energy and 
$\lambda=hc/\sqrt{T^2+2T E_0}$ the electron wavelength. 
The points where the electron wave enters and exits the spherical particle for a phase contour of radius $r_0$ are $\pm z_0(r_0)$, with
\beq
z_0(r_0)=\sqrt{R^2-r_0^2}
\eeq
with $R$ the radius of the particle. 
$V_0$ is the ordinary mean inner potential\cite{mip1,mip2} of the solid which
is expected to be constant inside the material. The additional potential $V_{ce}$  resulting from the charge redistribution 
shown in Fig. 3 is not constant but  depends  on the radial distance $r$  to the
center of the sphere:
\beq
r(z,r_0)=\sqrt{r_0^2+z^2} .
\eeq
The total potential energy of the electron wave inside the material
\beq
eV(z)=e(V_0 + V_{ce}(r(z,r_0)))
\eeq
depends on the position along the electron trajectory, as shown in Fig. 4 for several values of $r_0$. This additional potential well increases the phase shift of the high energy electron wave
going through the material, giving rise to the contour shifts shown in Figs. 1 and 2. 
For the examples of Figs. 1 and 2 we used $T=300keV$, $R=500 nm$.

\section{electric potential due to charge expulsion}

The theory of hole superconductivity\cite{hole1} predicts that the charge distribution in any superconductor at sufficiently low temperatures is macroscopically inhomogeneous, with 
more negative charge near the surface and more positive charge in the interior, resembling a `giant atom'\cite{giantatom}, as shown schematically in Figure 3. 
The charge inhomogeneity is very small: the excess negative charge resides within a London penetration depth (typically several hundreds Angstrom) of the surface and
is only of order 1 extra electron per  1 million atoms\cite{electrospin}. 

There is no direct experimental evidence of this physics so far, nor, we argue, is it ruled out by any existing experiment. Electron holography experiments on superconductors
performed in the past\cite{bonevich} have not tested this physics. A compelling reason in favor of this
scenario is that it provides a $dynamical$ explanation of the
Meissner effect\cite{sm}, not provided by the conventional BCS theory: in essence, magnetic field lines are dragged outward by outward electron flow.

The charge density distribution in the interior of a spherical superconductor of radius $R$ is predicted to be\cite{chargeexp}
\bmath
\beq
\rho(r)=\rho_0 [1-\frac{R^3}{3\lambda_L^2}\frac{1}{f(R/\lambda_L)}\frac{sinh(r/\lambda_L)}{r}]
\eeq
\beq
f(x)=xcosh(x)-sinh(x)  
\eeq
\beq
\rho_0=\frac{3E_m}{4\pi R} .
\eeq
\emath
$\rho_0$ is the uniform positive charge density deep in the interior of the superconductor.  The resulting electric field in the interior of the sphere is
\beq
\vec{E}(\vec{r})=\frac{E_m}{R}\vec{r}[1-\frac{R^3}{r^3}\frac{f(r/\lambda_L)}{f(R/\lambda_L)}]
\eeq
or, for $R>>\lambda_L$
\beq
\vec{E}(\vec{r})=\frac{E_m}{R}\vec{r}[1-e^{-(R-r)/\lambda_L}]
\eeq
giving the behavior shown qualitatively in Fig. 3. $E_m$ is the maximum value attained by the electric field near the surface in samples of radius much
larger than $\lambda_L$, and is given by
\beq
E_m=\frac{\hbar c}{4|e|\lambda_L^2}
\eeq
which is essentially the lower magnetic critical field $H_{c1}$ of a type II superconductor\cite{tinkham}. The resulting electric potential is
\beqn
V_{ce}(r)&=&\frac{E_m}{2R}(R^2-r^2)  \\ \nonumber
                 &+&\frac{E_mR^2}{f(R/\lambda_L)}[\frac{1}{r}sinh(\frac{r}{\lambda_L})-\frac{1}{R}sinh(\frac{R}{\lambda_L})]   .
\eeqn
The potential is always positive in the interior of the sample, of parabolic form except near the surface where it approaches zero with
vanishing slope. This positive potential enters in the expression for the phase shift Eq. (1) giving always an $increase$ in the phase shift
relative to the situation where the charge is uniform in the interior of the particle. The situation is similar to the case of a planar slab of material
discussed in Ref. \cite{holog1}.

\section{non-homogeneous samples}

For a perfectly homogeneous sample the theory predicts the situation shown in
Fig. 3. One may wonder whether the expected effect would be washed out in
non-homogeneous samples, because of  random positive and negative
phase shifts cancelling out.

In fact, that is $not$ what happens. Consider  a sample that has
random regions of superconducting material embedded in normal material,
with the superconducting regions of spherical shape for simplicity, as shown
in Fig. 5. As the high energy electron enters a superconducting region within
the sample 
it will always $speed$ $up$ first, decreasing its wavelength and hence
increasing its phase shift relative to the situation where the entire sample
is normal, and it will slow down to its original speed when exiting the
superconducting region. These increased phase shifts add up, {\it all with the
same sign}, as the electron travels through a non-homogeneous sample
with superconducting inclusions.

The reason that all the phase shifts are of the same sign is that the
superconducting regions of the sample $expel$ electrons. As a consequence,
the electric field inside the superconducting region points $towards$ the
nearest boundary surface to a normal region and acts to $accelerate$ the electron entering through
that surface. If instead
a superconducting region would have higher positive charge near its
boundary and higher negative charge in the interior, the beam electron
  traveling through that region would indeed
slow down and acquire a negative phase shift. However such a situation
$never$ happens within the theory discussed here.

 \begin{figure}
\resizebox{8.5cm}{!}{\includegraphics[width=7cm]{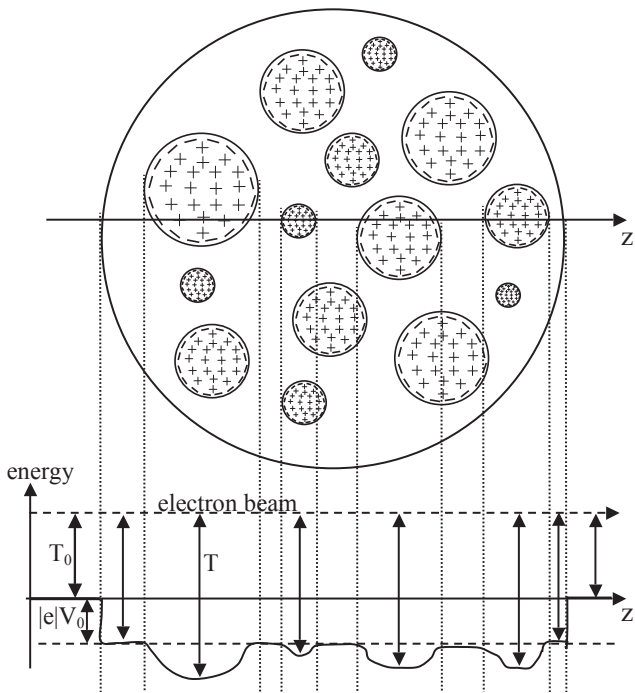}}
  \caption{ Schematic situation for a non-homogeneous sample,
  with superconducting and normal regions. The electron has kinetic energy $T_0$ outside the sample, and $T=T_0+|e|V_0$ inside the sample in the normal regions,
  with $V_0$ the normal state mean inner potential. As it
  enters a superconducting region it {\it speeds up}, its kinetic energy $T$
  (denoted by the vertical lines with double arrows) increases and its
  wavelength decreases, and as it exits the
  superconducting region it slows down and reverts to its original speed 
  and wavelength
  inside the normal sample. In the upper part of the figure the electron trajectory is denoted by the horizontal line labeled z, and the vertical dotted lines correlate the positions of the superconducting regions with the variations in the kinetic energy shown in the lower part of the figure.}
\end{figure} 

\section{temperature and material dependence}
The expressions for electric field and potential given in the previous section are the limiting values achieved at zero temperature where only the behavior of
the superconducting condensate is taken into account. At finite temperatures,
in a two-fluid description of the superconductor there will also be non-superconducting electrons, i.e. excited quasiparticles, that will tend to screen the
electric field produced by the charge redistribution of the condensate. Due to the existence of a superconducting energy gap, the density of 
quasiparticles goes to zero exponentially fast as the temperature is lowered. In Ref. \cite{holog1} we estimated that at temperature
$T\sim 0.1T_c$ the density of excited quasiparticles becomes sufficiently small that the electric field and electric potential resulting from charge expulsion
start to become visible. That estimate is likely to be an underestimate, since it ignored the fact that the charge of the excited quasiparticles is 
much smaller than the electron charge. In a more accurate calculation taking this fact into account we estimated recently that these effects should start
to become visible at temperatures $T\sim 0.16 T_c$\cite{explast}. 

However, we don't discount the possibility that small potential changes and resulting positive phase shifts may start to become observable 
immediately after the material is cooled below $T_c$,   i.e. at temperatures close to $T_c$, due to more complicated effects that have
not been taken into account in the existent theory. Even in the absence of macroscopic charge redistribution there should  be
$local$ charge rearrangement resulting from expansion of the electronic wavefunction in the superconducting state\cite{sm}, giving rise to a 
gradual increase in the mean inner potential as the material is cooled below $T_c$.
The observation of the phase shifts behaviors as a function of temperature in high
resolution electron holography experiments should play a key role in providing experimental input for further  development of the theory.

This theory is expected to applied to all superconductors, both the class termed ``conventional'' and the classes termed ``unconventional''. 
The properties of the particular superconducting material enter the predicted phase shifts through the values of $\lambda_L$ and $E_m$ (which is the same as $H_{c1}$). Smaller values
of $\lambda_L$ make the predicted effects larger, but they are usually associated with smaller values of $T_c$, putting larger demands on the experimental setup.
For the contours calculated in Figure 2, we assumed only two-times phase amplification and the parameters associated with the high temperature superconductor
$YBCO$, namely $\lambda_L\sim 150nm$ and $H_{c1}\sim 70G$. With a critical temperature of $T_c=90K$, these effects should be observable at temperatures of order
$10K$ and possibly substantially higher. All these conditions are well within the reach of current experimental facilities.

\section{discussion}

Electron holography provides a unique tool for the measurement of electrostatic potentials in the interior of materials, and thus is ideally suited for
examination of the physics discussed here. The spherical samples discussed in this paper should make the interpretation of the observed results
particularly simple because of the absence of electric fields outside the sample, but the predicted effect should also be seen for 
more general shapes. For non-spherical shapes, electric fields outside  the sample are 
also predicted to be generated as a consequence of  charge expulsion\cite{ellipsoid}, with electric field lines going from regions of low suface curvature to regions of
high surface curvature. In a subsequent step it will be interesting to use electron holography 
to study  such samples and detect electric potential variations both inside and outside the sample to  confront with the theoretical predictions.

Admittedly, the examples of phase shifts shown in Figs. 1 and 2  are  for sample thicknesses somewhat larger than typically used in electron holography
experiments (a few hundred nm). As the sample thickness increases one has to worry about   effects such as
(i) inelastic scattering processes that would increase the tempearature of the sample, and (ii) generation of secondary electrons and resulting
positive charging of the sample. On (i) we note that it can be minimized by reducing the intensity of the electron beam. On (ii) we
point out that any such charging effect would presumably occur both above and below $T_c$ so  that one can still hope to discern the effect
discussed here specifically due to superconductivity.
We also point out that the examples shown in Figs. 1 and 2, with  particle radii 500 nm, assumed only two-times phase amplification. With phase shift sensitivity of
e.g. $2\pi/100$\cite{endo2} one would detect an effect for the superconducting material parameters assumed in Figs. 1 and 2 for
particle radii as small as  approximately  130 nm and 250 nm respectively.

With the high spatial resolution currently attainable in electron holography experiments it should be possible to measure the magnitude of the shift of a contour line
as function of the radius of the contour line very accurately, and contrast it with the predictions resulting from our theory, Eqs. (1), (4), (5) and (10). It should 
also be of great interest to obtain spatial maps of
phase shifts, potentials and charge densities for non-homogeneous samples, and in particular to study the temperature dependence
of these quantities in the neighborhood of grain boundaries\cite{grain}.

Another prediction of this theory that may be amenable to test by electron holography is that electrons will ``spill out'' from the surface of the sample as the temperature is lowered
below $T_c$\cite{giantatom}. This may have an observable effect on Fresnel fringes arising from
diffraction from  the edge of a superconducting sample and will be explored in future work.

The apparent increase in thickness referred to in the title of this paper is of course not a real change in thickness but a virtual one, 
a pictorial way of describing the increase in the mean inner   potential
predicted to take place in superconductors at low temperatures. Alternatively, one may describe the predicted effect as an
apparent increase of the atomic number of the ions in  the  material, $Z$. It is interesting to note that the theory predicts the largest tendency for superconductivity
for materials where conduction occurs through negatively charged ions\cite{materials}, hence
the effective ionic charge $Z$ sensed by the conduction electrons in the normal state is smallest, and  the excess negative charge
in the conducting substructures is largest. As the sample becomes superconducting it attempts to counter these effects.

It is also interesting to note that the mean inner potential is related to the diamagnetic susceptibility of a material\cite{diamag}: materials with higher mean inner potential have higher diamagnetic susceptibility\cite{diamag,diamag2,diamag3}. As a system goes superconducting its diamagnetic susceptibility of course  increases dramatically, and thus it 
could be argued that it is   not surprising that its mean inner potential should increase, as predicted here. 
However, conventional BCS theory of superconductivity does $not$ predict such an effect.
The relation between diamagnetic susceptibility and mean inner potential follows from the fact that both can be shown to be proportional to the mean
square atomic radius of the electronic wavefunction\cite{bethe,mip1}. Within our theory, superconductivity is associated with $expansion$ of the electronic wavefunction\cite{explast,meissner}, and thus it is natural that it is associated with an increase in the mean inner potential. This is not the case in BCS theory.

Or, put another way: as discussed in ref.\cite{mip1}, the mean inner potential ``has two important interpretations - first, as a measure of diamagnetic susceptibility, and
secondly, as a measure of the `size' of an atom''. Within our theory, a superconductor is a `giant atom' because of its magnetic $and$ electric properties that
include charge redistribution\cite{giantatom}, and the connection remarked by Spence\cite{mip1} qualitatively holds. In the conventional theory of superconductivity superconductors have
also   been described as `giant atoms' by London\cite{london} and others\cite{london2,london3} but only with respect to  their magnetic properties, hence no implication for the mean inner potential results from it and the
connection between mean inner potential and diamagnetic susceptibility does not exist.
The fact that within our theory the connection between diamagnetic susceptibility and mean inner potential does exist strongly suggests that small positive phase shifts
will onset immediately below $T_c$, as the diamagnetic susceptibility starts to increase due to wavefunction expansion\cite{meissner}. 
However only far below $T_c$ should the larger phase shifts resulting from macroscopic charge redistribution discussed here appear.

Note that when  the temperature of a material is lowered, its dimensions usually decrease rather than increase. A key contrary example is the behavior of superfluid $^4He$, which $expands$ as the temperature is lowered below the $\lambda$ transition. We have proposed elsewhere\cite{helium,helium2} that  
the negative thermal expansion of $^4He$ in its superfluid state is intimately related to the physics discussed here for superconductors, and results from the fact that both the transitions to   the superfluid $^4He$ state and to  the
superconducting state of metals are driven by {\it lowering of kinetic energy}\cite{kinetic,meissner},
or equivalently {\it quantum pressure}\cite{emf}. Still another example of this physics is the 
anomalously small thermal expansion observed in metallic ferromagnets\cite{ferrothermal}, which can also be understood if metallic ferromagnetism is   driven
by lowering of kinetic energy\cite{ferro}.

As we have discussed for homogeneous as well as 
non-homogeneous samples, the predicted
phase shifts for electrons traveling through superconducting regions
are always positive. In other words, the beam electron's kinetic energy 
increases. This fact reflects the fundamental
electron-hole asymmetry on which the theory of superconductivity
discussed here is based\cite{ehasym}: if positive rather than negative charge
would sometimes be expelled from 
the interior of superconducting regions towards the surface, the 
phase shifts for beam electrons traveling through those regions would
be negative. That will $never$ happen according to our theory.
It is interesting that the kinetic energy $decrease$ of the superconducting
electrons predicted by our theory\cite{kinetic} that is associated with the negative charge
expulsion is mirrored by the kinetic energy $increase$ of the
beam  electrons traveling through the sample   predicted to take place
 in  electron holography experiments.

The physics underlying the properties of the superconducting state discussed here is   expansion of the electronic wave function
and associated expansion of the negative charge cloud as
phase coherence is established through the superconducting
regions of the  sample, originating in increased outward quantum pressure.
It is very  remarkable  that the outward motion of the constant phase contours imaged by electron holography will provide a vivid picture
of this expansion physics driven by quantum pressure, since it mimics the behavior that would result if the radius of the sample was  increasing, even though the
physical radius of the sample is actually not increasing. It will be, in the words of A. Tonomura,  a new demonstration of ``the quantum world unveiled by electron waves''\cite{quantumworld}.

\acknowledgements
The author is grateful to J.C.H. Spence, R. Dunin-Borkowski, H. Lichte, C.T. Koch, E. Voelkl and Y. Zhu for stimulating  discussions.

 \end{document}